\documentclass[smallcondensed]{svjour3}
\pdfoutput=1
\usepackage{multirow}
\usepackage{color} 
\usepackage{caption}
\usepackage{graphics} 
\usepackage{rotating}
\usepackage{eqparbox}
\usepackage{graphics}
\usepackage{colortbl} 
 \usepackage{mathptmx} \usepackage[scaled=.90]{helvet} \usepackage{courier}
\usepackage{balance}
\usepackage{picture}
\usepackage{algorithm}
\usepackage{algorithmicx}
\usepackage{algpseudocode}
\usepackage[export]{adjustbox}
\usepackage{amsmath}
\usepackage{listings}
\usepackage{mdframed}
\usepackage{rotating}
\definecolor{Gray}{rgb}{0.88,1,1}
\definecolor{Gray}{gray}{0.85}
\definecolor{Blue}{RGB}{0,29,193}

\definecolor{lightgray}{gray}{0.8}
\definecolor{darkgray}{gray}{0.6}

\definecolor{Gray}{gray}{0.95}
\definecolor{LightGray}{gray}{0.975}

\newcommand{\etal}{et al.}

\newcommand{\bi}{\begin{itemize}[leftmargin=0.4cm]}
\newcommand{\ei}{\end{itemize}}
\newcommand{\be}{\begin{enumerate}}
\newcommand{\ee}{\end{enumerate}}
\newcommand{\tion}[1]{\S\ref{sect:#1}}
\newcommand{\fig}[1]{Figure~\ref{fig:#1}}
\newcommand{\tab}[1]{Table~\ref{tab:#1}}
\newcommand{\eq}[1]{Equation~\ref{eq:#1}}

\usepackage{soul}

\usepackage[shortlabels]{enumitem}  
\usepackage{url}
\usepackage{graphicx}
\usepackage{multirow}
\usepackage[svgnames]{xcolor}
\usepackage[framed]{ntheorem}
\usepackage{framed}
\usepackage{tikz}
\usetikzlibrary{shadows}
\theoremclass{Lesson}
\theoremstyle{break}

\tikzstyle{thmbox} = [rectangle, rounded corners, draw=black,
  fill=Gray!20,  drop shadow={fill=black, opacity=1}]

\newshadedtheorem{lesson0}{Result}
\newshadedtheorem{lesson}{Result}

\usepackage{xcolor}
\usepackage{lipsum}

\begin{document}
\date{}
\title{ Impacts of Bad ESP (Early Size Predictions) on Software Effort Estimation}
\author{George Mathew \and Tim Menzies \and Jairus Hihn}
\institute{
    G. Mathew, T. Menzies \at
                Department of Computer Science, North Carolina State University, Raleigh, NC, USA\\
                \email{george.meg91@gmail.com, tim.menzies@gmail.com}
    \and
    J. Hihn \at Jet Propulsion Laboratory, Pasadena, CA, USA \\ \email{jairus.m.hihn@jpl.nasa.gov}}
% \ead{george.meg91@gmail.com}
% \author[add1]{}
% \ead{tim.menzies@gmail.com}
% \author[add2]{Jairus Hihn}
% \ead{ jairus.m.hihn@jpl.nasa.gov}
% \cortext[cor1]{Corresponding author: Tel:1-614-535-8678(George)}
% \address[add1]{Department of Computer Science, North Carolina State University, Raleigh, NC, USA}
% \address[add2]{Jet Propulsion Laboratory, Pasadena, CA, USA}

\small
% % \numberofauthors{1}
% \author{Wei Fu \and Tim Menzies \and Xipeng Shen}
% \institute{North Carolina State University, Raleigh, NC, USA
%       Wei Fu \email{w}}
% % \email{fuwei.ee \and tim.menzies@gmail.com \and xshen5@ncsu.edu }

% \thispagestyle{plain}
% \pagestyle{plain}
\maketitle
\vspace{-1in}
\begin{abstract}
  {\em Context:}
  %For  large systems (e.g.  projects run by government or
%defense departments), it is common to lobby for the development funds prior to commencing the work.
%For such software systems, it is important to have an (approximately) accurate    early life cycle effort estimate % since (1)~large sums of money are involved and (2)~once funds are allocated, it can be problematic
%to lobby for further funds. However, before the software is built, the size of the final system is not
  %known.
  Early size predictions (ESP) can lead to errors in effort predictions for software projects.
  This problem is particular acute in parametric effort models that give extra weight to
  size factors (for example, the COCOMO model assumes that effort is exponentially proportional
  to project size).\\
  {\em Objective:} To test if effort estimates are crippled by bad ESP.
   \\
  {\em Method:} Document  inaccuracies in early
  size estimates. Use those error sizes to determine the implications of
  those inaccuracies via an Monte Carlo perturbation analysis of effort models
  and an analysis of the equations used in those effort
  models.\\
{\em Results:} 
While many 
projects have errors in ESP of up to $\pm$~100\%,
those errors add very
little to the overall effort estimate error. Specifically,
we find no statistically significant difference in
the estimation errors seen after
increasing ESP errors from 0 to $\pm$~100\%.
An analysis of effort estimation models explains why this is so:
the net additional impact  of ESP error is relatively  small compared to the other sources of error associated within
estimation models.\\
{\em Conclusion:} ESP errors effect effort estimates by a relatively
minor amount. As soon as a
model  uses a size
estimate {\em and other factors} to predict project effort,
then ESP errors  are not crippling to the process of estimation.
\end{abstract}

% % A category with the (minimum) three required fields
% \vspace{1mm}
% \noindent
% {\bf Categories/Subject Descriptors:} 
% D.2.8 [Software Engineering]: Product metrics;
% I.2.6 [Artificial Intelligence]: Induction

\vspace{1mm}
\noindent
{\bf Categories/Subject Descriptors:} 
D.2.9 [Software Engineering]: Time Estimation;
K.6.3 [Software Management]: Software Process

\vspace{1mm}
\noindent
{\bf Keywords:} software effort, parametric models, COCOMO, bootstrap sampling, effect size
%  \maketitle 
%\pagenumbering{arabic} %XXX delete before submission

\section{Introduction}

Poor software effort estimation can cripple a project.
In
the worst case, over-running projects are canceled and
the entire development effort is wasted.
One challenge with   effort estimates
is  {\em Early Size Prediction} (ESP) where   size
estimates are generated  early in the project life cycle. For large projects, or projects
with extensive government or regulatory oversight, such estimates are needed for many purposes
including   justifying cost for different design decisions; 
debating which projects get funded or not;
when  a  project is delivered late (or not at all) and when the management's initial decisions are being audited.

Incorrect size estimates can have a large impact on effort estimates. 
Boehm~\cite{boehm81} proposes a parametric model of effort estimation that
  assumes size was proportional exponentially to effort; i.e.
  \begin{equation}\label{eq:one}
    \mathit{effort} = \mathit{X*KLOC}^Y
    \end{equation}
  where {\em KLOC} is thousands of lines of code and $X,Y$ represent numerous ``context variables'' such as analyst experience.
  \eq{one} is the basis for the COCOMO model~\cite{boehm81,boehm00b}  (described in detail later in this paper).
  In this equation, ESP errors in KLOC can lead to exponentially large errors
  in the  effort estimate.
  This is a concern since it may be very difficult to estimate KLOC
  early in the software development process.
  For example, consider the  issues that needs to be resolved in order
  to make a highly accurate size estimate.
  How was reused code counted in the size estimate?
  Was size measured from end-statement or end--line?
 How were lines of comments handled?
 How to estimate the size of systems written in multiple languages? 
  How to guess early in the life cycle
  how much subsequent  feedback will change
  the goals of the project and the size of the delivered system?
Given the open-ended nature of all the above, it is seems reasonable to conclude that ESP errors will cripple any size-based estimation process.

 This paper argues for another  conclusion. While there are many  {\em potential} problems with size estimation,
 we show  that the actual
  net effect of ESP errors is surprisingly small. {\em Empirically}, we use a sensitivity analysis to show that    random perturbations of the input values to \eq{one} lead to  remarkable
little error in the overall estimate.
Further, we can explain this 
  {\em analytically}   by  deriving  the ratio of the {\em least} to {\em largest} estimate within \eq{one}; i.e. $\frac{\mathit{largest}}{\mathit{least}}=2028*\mathit{KLOC}^{0.74}$. 
Note the   small coefficient on lines of code (0.74) and the   large linear coefficient showing the effects of all other parameters (2048).
What this expression shows is that decisions about the other parameters can be lead to much larger estimation variance than decisions just about KLOC.
Hence, we assert that:
\begin{quote}
{\em The impact of bad ESP  may be 
 relatively less important than other factors.} 
\end{quote}
Our new conclusion is defended in  detail in section \ref{sect:rqs} using  three research questions:
\bi
    \item \textbf{RQ1}: How big are real world ESP errors?
    \item \textbf{RQ2}: What is the impact of real-world ESP errors?
    \item \textbf{RQ3}: Within an effort  model, what is the maximum effect of making large changes to a size estimate?
\ei
The rest of this paper  describes studies that motivate this work and
the general theory of  effort estimation. After that, we 
   present the data and experimental methods
used in this study.  Using that data and methods, our three research
questions are then explored.

\subsection{Notes}
Before beginning, we digress to make some important points.
Firstly, this paper uses an old effort model (COCOMO) to explore these ESP issues. Why? Surely there are more recent, better, alternatives?

It turns out that, even after decades of work on other approaches, COCOMO remains remarkably more effective than other models. Recently~\cite{Menzies2016},
we compared COCOMO against the current state of the art (reasoning via analogy using spectral-based clustering plus instance and feature selection, and a recent "baseline method" proposed in ACM Transactions on Software Engineering~\cite{Whigham:2015}).
Much to our surprise, the estimates from old COCOMO were as good, or better, than those from the more recent methods.
This is not to say that the newer methods are useless since not all project
data can be expressed using the terminology acceptable to COCOMO.
However, what that does say, is that it is remains valid to use COCOMO as a workbench where we will explore
bad ESP effects.  
For further notes on why COCOMO is worthwhile, useful and relevant  to  study, see \tion{altCoc}.

Secondly, it should be stressed
that the goal of this paper is {\em not} to propose a better method for effort
estimation.  Rather, our goal is to is to say that an issue at the heart of all
estimation (guessing the properties of something before that something is built)
is not a major problem. This has several implications:

\bi
\item {\em Practical implications \#1:} This paper removes
  a major objection to the use of effort estimation models.
\item {\em Practical implications \#2:} Dozens of the projects studied in this paper come from very speculative systems (NASA flight systems) or incremental
  systems where it is most likely that ESP will be inaccurate. Yet this paper shows that the net effect
  of to inaccuracies is very small.
  Accordingly, we say that
  the results of this paper demonstrate
  that software engineers could make more use of effort estimation even when exploring incremental or highly experimental methods. 
  \item {\em Theoretical implications:} This paper offers a methodology for testing the impact
  of size inaccuracies on effort estimation models. In terms of 
  future publications that cite this work, we anticipate that that 
  methodology will be the most widely used part of this paper.
  \item {\em Methodological implications:}
   Numerous
  recent publications caution that merely because some  belief that are widely held by
  supposed  experts, they can still be misleading~\cite{jorgensen09,Menzies2016,Menzies2016b,passos11,prem16,betten14,yang13,me12d,ray2014lang}.  
   All the evidence required to make the analysis of
  this paper has been available since  2010-- yet in all that
  time no one has thought to use it to test the
  SE truism that  ESP can be very problematic.
  We hope that this paper inspires other researchers to revisit
  old truisms in this field (which may not be outdated).
  \ei

  \section{Motivation}
  The analytical analysis of this paper were motivated by some recent very
  curious  results based on a {\em sensitivity/uncertainty analysis} of  effort estimation models~\cite{spip09}.
  This section describes those prior results.
  
  Sensitivity analysis (SA) refers to a suite of  
 techniques   that determine how different values of model independent variables
 (a.k.a. model inputs) impact a particular dependent variable (a.k.a. model output) under a given set of assumptions~\cite{sal00}. 
 Sometimes an  SA is combined with an uncertainty analysis (UA) that seeks to quantify the uncertainty to (say)
 determine how much uncertainty on the input space can be tolerated before losing the ability to make
 precise statements about the model.   
 
For simple models that are continually differential at all parts of the internal space, SA/UA can be
conducted   by ``surfing the surface''; i.e. move in the direction of most change
(as shown by 
  partial differentials computed within the model), then  reporting the effects on the output variables~\cite{gay2010automatically}.
But some models   are not continually differential. For example, XOMO~\cite{Menzies:2007men} is a combination
of the  COCOMO effort models plus other models that predict for project risk and defects.
XOMO   contains numerous
if-statements that divide the internal space of the model into discontinuous sub-regions.
Non-parametric SA/UA  can be performed on  such non-differential models using a biased sampling approach.
Many thousands of times (or more), inputs are drawn from known input distributions and the model is executed. 
XOMO's supports UA by representing the internal parameters of COCOMO not as points values  but as ranges (so each time
the model is run, different values may be drawn across the ranges).

In 2009, with Boehm, Madachy et al.~\cite{spip09}, we  reported a most curious aspect of the SA/US results from XOMO. In the study, the $E_1$ errors seen from using normal COCOMO (with no uncertainty analysis)
were compared to the $E_2$ errors seen in XOMO's UA estimates. In that comparison,  
the $E_2$ effort estimation errors were not much larger than $E_1$  (an additional 20 to 25\%, relative to $E_1$).  
This is a most remarkable result: the $E_2$ estimation errors are comparatively very small even though they
 were  generated from an extensive uncertainty analysis that   
included large  KLOC perturbation (e.g. in those case study, KSLOC was perturbed by a factor of 1.5 to 10).

In 2012, another study reported  that the effort estimate errors are remarkably insensitive
to perturbations in KLOC. Kocagunelli et.al.~\cite{Kocaguneli2012z} built effort estimates with and without
size data. They found that (a)~removing size features decreased the performance; but (b)~during an extensive sensitivity analysis by perturbing random features(other than size), they achieved surprisingly effective predictions.
Based on these results a natural question arises on the effect of uncertainty in size estimates on parametric effort models.

Prior to the analysis of this current paper, it was inexplicable
why   changing~\cite{spip09} or ignoring~\cite{Kocaguneli2012z}  size
estimates results in  only small changes in the final estimate.
However, given the analytical results described below, this is no longer the case since
(as stated in the introduction) decisions about other   things  can  be  lead  to  much  larger  estimation variance than decisions just about KLOC. 
The rest of this paper presents (a)~some background theory on effort estimation;
(b)~one more empirical study that confirms the two results shown here (that 
the impact of real-world ESP errors are remarkably small) and (c)~derives analytically
a mathematical expression showing we should expect ESP errors to have relatively little overall effect.

  \section{Effort Estimation: Theory}

\subsection{On the Importance of Size Estimates}

Many  authors have discussed issues related to size estimates in effort
estimation.  One class of comment is that measuring size is a meaningless
measure and we just should not do it.  Quoting the former CEO of Microsoft, Bill
Gates~\cite{goll04}:
\begin{quote}{\em ``Measuring software productivity by lines of code is
  like measuring progress on an airplane by how much it weighs.''}\end{quote}
A similar
complaint was made by Dijkstra~\cite{dij88} who says\begin{quote}
{\em ``This (referring to
  measuring productivity through KLOC) is a very costly measuring unit because it
  encourages the writing of insipid code, but today I am less interested in how
  foolish a unit it is from even a pure business point of view.''}
  \end{quote}
On the other hand, several studies report that size estimates have a place in
effort estimation:
\bi
\item
Walkerden and Jeffery comment that estimating effort via linear size
adjustment to an analogue is more accurate than estimates based on regression
model~\cite{Walkerden1999}.
\item
  Kirsopp \etal agree with Walkerden and Jeffery: in a follow-up paper
they noted that  linear size scaling adaptation results in statistically
significant improvements in predicting
effort~\cite{kirsopp2003empirical}.
\item  J{\o}rgensen  et al. observe several industrial software development and
maintenance projects. They note that  that the effort estimates provided by software
professionals (i.e. expert estimates) are to some extent are based on adjustments
to size by regressing towards the mean(RTM).~\cite{jorgensen2003software}. Which is to say
that the predictions of human estimators are also Thus, an error in estimating the size
of the project can lead to drastic change in predicting the effort of the
project.
\ei
This paper offers a  middle ground between those who claim size estimates are irrelevant and those
who say they are fundamental to the estimation process.
 Like Walkerden, Jeffery, Kirsopp, and J{\o}rgensen  et al., we say that size estimates matter somewhat.
 However, as suggested by Gates and Dijkstra, size estimates are   not vitally important to effort estimation
 since effort error comes from two factors:
 \bi
\item Uncertainty in the size estimate;
\item Uncertainty in all the other project factors used in the estimation models;
  \ei
  and our  results show that
  uncertainties in the size estimates effect the effort estimate much less than uncertainties in the other project factors.
  
% Based on numerous studies, size of the project is considered the most predictable feature that quantifies the effort involved in a developing a project and has the largest correlation with it.~\cite{Walkerden1999, kirsopp2003empirical, jorgensen2003software}. Thus, an error in estimating the size of the project can lead to drastic change in predicting the effort of the project. For example, Tronto \etal \cite{de2007comparison} recommended the use of multi-layered Artificial Neural Networks for estimating Software effort. It is well known that the weights of the neural network can change drastically on the smallest change in the independent variables. In other studies by Huang \etal~\cite{huang2006optimization} Genetic Algorithm based Analogy based methods also depend on the size of the project and an error in the estimation of size can lead to false optimization of software effort. Checking project similarities also heavily depend on size of the project and in clustering based methods like C-means~\cite{azzeh2008software} a change in the size estimate can change cluster associated with the project.

  \subsection{Models of Effort Estimation}\label{sect:emodels}
  There are many models of effort estimation.
   For example, Shepperd et
 al. prefer non-parametric analogy-based estimation (ABE)
 methods~\cite{shepperd1997estimating}.  Meanwhile, within the
 United States Department of Defense and NASA,
 parametric effort models like
 COCOMO~\cite{boehm81} are used extensively and
 have found to be quite effective~\cite{lum02}.
 Also, advocates of agile and devops methods prefer
 the planning poker method (discussed in \tion{altCoc}).
 Finally J{\o}rgensen  et al.~\cite{jorgensen09} prefer
 expert-based approaches where estimates are derived
 by committees of human experts.  For studies
 comparing these methods, or how to usefully combine
 them, and how to reduce errors when they are used,
 see~\cite{koc11b,Minku2013,garg15,me13a}.

 \begin{table}[!t]
{\scriptsize
\begin{center}
\begin{tabular}{|c|p{0.9in}|p{0.9in}|p{0.9in}|p{0.9in}|}\hline

 & Definition & Low-end = \{1,2\}
 &Medium =\{3,4\} &High-end= \{5,6\} \\\hline

\multicolumn{1}{c}{~}\\

\multicolumn{5}{l}{\textbf{Scale factors(SF)}}\\\hline
Flex   &  development flexibility   & development process
rigorously defined & some guidelines, which can be relaxed & only
general goals defined\\\hline

Pmat    & process maturity  &  CMM level 1 &   CMM level 3  &  CMM level 5 \\\hline

Prec & precedentedness  &  we have never built this kind
of software before &    somewhat new &
thoroughly familiar \\\hline

Resl &  architecture or risk resolution  &  few interfaces
defined or few risks eliminated  &  most interfaces defined or most
risks eliminated   & all interfaces defined or all risks
eliminated\\\hline

Team  &   team cohesion  &  very difficult interactions &
basically co-operative  &  seamless interactions\\\hline

\multicolumn{1}{c}{~}\\

\multicolumn{5}{l}{\textbf{Effort multipliers(EM)}}\\\hline
acap  &  analyst capability  &  worst 35\% &   35\% - 90\% &  best 10\% \\\hline

aexp   &  applications experience  &  2 months &   1 year  &  6 years\\\hline

cplx   &  product complexity   & e.g. simple read/write
statements & e.g. use of simple interface widgets  &  e.g.
performance-critical embedded systems\\\hline

data   &  database size 
(DB bytes/SLOC) &
10 & 100 &    1000 \\\hline

docu   &  documentation   & many life-cycle phases not
documented      & &  extensive reporting for each life-cycle phase\\\hline

ltex   &  language and tool-set experience   & 2 months  &  1
year & 6 years \\\hline

pcap   &  programmer capability  &  worst 15\%   & 55\%  &  best 10\% \\\hline

pcon   &  personnel continuity \newline
(\% turnover per year) &
    48\% &    12\%  & 3\% \\\hline

plex   &  platform experience  &  2 months  &  1 year  &  6 years\\\hline

pvol   &  platform volatility ($\frac{frequency~of~major~changes}{frequency~of~minor~changes}$) &
$\frac{12~months}{1~month}$   & $\frac{6~months}{2~weeks}$ &
$\frac{2~weeks}{2~days}$\\\hline

rely   &  required
reliability &   errors are slight inconvenience  &  errors are easily
recoverable   & errors can risk human life\\\hline

ruse   &  required
reuse &   none &    multiple program  & multiple product lines\\\hline

sced  &   dictated development\newline schedule &    deadlines moved to
75\% of the original estimate &  no change
&  deadlines moved back to  160\% of original estimate\\\hline

site   &  multi-site development   & some contact: phone, mail&
some email  &  interactive multi-media\\\hline

stor  &   required \% of available
RAM & N/A
 &   50\% &  95\% \\\hline

time  &   required \% of available CPU &
N/A&     50\%
   &  95\% \\\hline

tool   &  use of software tools  &  edit,code,debug &&
integrated with life cycle\\\hline

\multicolumn{1}{c}{~}\\

\multicolumn{5}{l}{\textbf{Effort}}\\\hline

months & construction effort  in months& \multicolumn{3}{l|}{1 month =  152 hours (includes development \& management
hours).  
}\\\hline
\end{tabular}
\end{center}
} \caption{COCOMO-II attributes.}
\label{tab:cparems}
\end{table}

As to the effect of bad ESP on human-generated estimates, we repeat the comments of
J{\o}rgensen and Gruschke~\cite{jorgensen09}: the best way to reduce error
in human-generated estimates is to not fix bad ESP but to
conduct lessons-learned sessions after each project. 
 
As to the effects of bad ESP on  non-parametric models, it is
 trivially simple to show that bad ESP only has minimal effect on ABE.
 ABE generates estimates via a distance metric
 defined over all $F$ factors that typically includes one size estimate\footnote{Evidence:
   in the 13  effort data sets of the PROMISE repository http://openscience.us/repo/effort, only one data set has more than a single
   size attribute.}. ABE does not give extra special weight to the size factor.
 On the contrary, ABE often mutes the impact of the size factor.
The range of values in a size factor may be much larger than all the other factors; e.g. programmer
 capability may be scored on a six point integer scale while the estimates may
 range from zero to millions of lines of code. When faced with scales of very
 different sizes, standard practice~\cite{koc11b,aha1991instance} is to
 normalize all the numerics min to max, zero to one using
 \mbox{$(x-\mathit{min})/(\mathit{max} - \mathit{min})$}.

  Note that such normalization is recommended practice. 
 Kocaguneli et al.~\cite{Kocaguneli2012z}, studied the effects of  normalization
 within 90 varieties of effort estimation. That study found 13 ``superior''
 methods and 77 ``inferior'' ones. Of the eight varieties that never normalized,
 seven were found in the ``inferior'' set.

 The point here is that once the size factor is muted via normalization,
  then ABE estimates are effected by bad ESP by a factor of
   $1/F$. Typically values for $F$ are 5 to 24
  with medians of around ten. That is, when compared to the other $1-1/F$ factors,  bad ESP has relatively
  little impact on the ABE effort estimate.
 
% \new{A similar argument can be made for other non-parametric methods such as ensemble
% methods~\cite{Kocaguneli2012z} or random forests~\cite{breiman2001random} of
% CART regression trees~\cite{breiman1984classification}. When those models contain
% $F$ factors only one of which is a size estimate, and those models make no
% special use of size, then here exists $1-1/F$ other factors that contribute more
% to effort estimation error. Menzies \etal strengthens this argument when in 2009 they empirically show from 10 different case studies that the uncertainty of parameters in effort estimation models are more dominant compared to the uncertainty of contribution associated with the parameters~\cite{menzies2009accurate}}

The problem of bad ESP is most acute for parametric models particularly those such as \eq{one} that give
extra special weight to the size estimates.
Such parametric models are widely used, particularly for large government project.
In our work with the Chinese and the United States software industry,
we see an   almost exclusive
use  of parametric estimation tools such as those offered by 
Price Systems (pricesystems.com) and  Galorath (galorath.com).
Also,
professional societies, handbooks and
certification programs are mostly developed around 
parametric estimation methods and tools; e.g. see the 
International Cost Estimation and Analysis Society; the
NASA Cost Symposium;  the
International Forum on COCOMO and Systems/Software
Cost Modeling\footnote{See the web sites \url{http://tiny.cc/iceaa}, \url{http://tiny.cc/nasa_cost}, \url{http://tiny.cc/csse}}.

 This paper uses the COCOMO model as a representative of the parametric models since it is open source.
 As to other parametric effort models,
 Madachy and Boehm~\cite{madachy2008comparative} report that many aspects of this model
 are shared by other models in widespread commercial
 use such as  SEER-SEM~\cite{boehm00b} and Price-S (now called True S).

 \begin{figure}[!t]
\begin{lstlisting}
_  = None;  Coc2tunings = [[
#              vlow  low   nom   high  vhigh  xhigh   
# scale factors:
'Flex',        5.07, 4.05, 3.04, 2.03, 1.01,     _],[
'Pmat',        7.80, 6.24, 4.68, 3.12, 1.56,     _],[
'Prec',        6.20, 4.96, 3.72, 2.48, 1.24,     _],[
'Resl',        7.07, 5.65, 4.24, 2.83, 1.41,     _],[
'Team',        5.48, 4.38, 3.29, 2.19, 1.01,     _],[
# effort multipliers:        
'acap',        1.42, 1.19, 1.00, 0.85, 0.71,    _],[
'aexp',        1.22, 1.10, 1.00, 0.88, 0.81,    _],[
'cplx',        0.73, 0.87, 1.00, 1.17, 1.34, 1.74],[
'data',           _, 0.90, 1.00, 1.14, 1.28,    _],[
'docu',        0.81, 0.91, 1.00, 1.11, 1.23,    _],[
'ltex',        1.20, 1.09, 1.00, 0.91, 0.84,    _],[
'pcap',        1.34, 1.15, 1.00, 0.88, 0.76,    _],[ 
'pcon',        1.29, 1.12, 1.00, 0.90, 0.81,    _],[
'plex',        1.19, 1.09, 1.00, 0.91, 0.85,    _],[ 
'pvol',           _, 0.87, 1.00, 1.15, 1.30,    _],[
'rely',        0.82, 0.92, 1.00, 1.10, 1.26,    _],[
'ruse',           _, 0.95, 1.00, 1.07, 1.15, 1.24],[
'sced',        1.43, 1.14, 1.00, 1.00, 1.00,    _],[ 
'site',        1.22, 1.09, 1.00, 0.93, 0.86, 0.80],[ 
'stor',           _,    _, 1.00, 1.05, 1.17, 1.46],[
'time',           _,    _, 1.00, 1.11, 1.29, 1.63],[
'tool',        1.17, 1.09, 1.00, 0.90, 0.78,    _]]

def COCOMO2(project,  a = 2.94, b = 0.91, # defaults
                      tunes= Coc2tunings):# defaults 
  sfs ems, kloc  = 0,1,22          
  scaleFactors, effortMultipliers = 5, 17
  for i in range(scaleFactors):
    sfs += tunes[i][project[i]]
  for i in range(effortMultipliers):
    j = i + scaleFactors
    ems *= tunes[j][project[j]] 
  return a * ems * project[kloc] ** (b + 0.01*sfs) 
\end{lstlisting}
\caption{COCOMO-II: effort estimates from a {\em project}.
Here, {\em project} has up to 24 attributes  (5 scale
factors plus 17 effort multipliers plus KLOC plus. in the training data, the actual effort).
Each attribute except KLOC and effort is scored
using the scale very low = 1, low=2, etc.
For an explanation of the attributes shown in
green, see \tab{cparems}.}\label{fig:coc2}
\end{figure}

  \subsection{COCOMO Details}

  COCOMO was developed in two states: an initial release in 1981~\cite{boehm81}
  followed by an extensive revision in  2000~\cite{boehm00b}.
  In between those releases,
  Boehm created a consortium for
industrial users of COCOMO.
This consortium
collected information on 161 projects from commercial,
aerospace, government, and non-profit organizations.
Using that new data, in 2000, Boehm and his colleagues developed
a set of   {\em tunings} for COCOMO-II that
mapped the project descriptors (very low, low, etc)
into the specific attributes used in the COCOMO model (see \tab{cparems}).
Those tunings, mappings, and attributes became the COCOMO-II model
released 
\begin{equation}\label{eq:cocII}
\mathit{effort}=a\prod_i EM_i *\mathit{KLOC}^{b+0.01\sum_j SF_j}
\end{equation}
Here, {\em EM,SF} are effort multipliers and scale factors and $a,b$ are the
{\em local calibration} parameters (with default values of 2.94 and 0.91).  In
COCOMO-II, effort multipliers change effort by a linear amount while scale
factors change effort by an exponential amount.  COCOMO-II reports {\em effort}
as ``development months'' where one month is 152 hours of work (and includes
development and management hours).  For example, if {\em effort}=100, then
according to COCOMO, five developers would finish the project in 20 months.

For a complete implementation of the COCOMO-II effort model, see \fig{coc2}.
Note that \eq{cocII} defines the internal details of $X,Y$ terms in \eq{one}:
\mbox{$X=a\prod_i EM_i$} and \mbox{$Y=b+0.01\sum_j SF_j$}.

\subsection{Alternatives to COCOMO} \label{sect:altCoc}

% \noindent\colorbox{lightgray}{%
%     \parbox{\dimexpr\linewidth-2\fboxsep}% a box with line-breaks that's just wide enough
%         {Note to reviewers: we are not sure if this paper requires this section. None of the material here is needed for the rest of the paper. Your comments on this matter would be appreciated.
%         }
% }
 
COCOMO was initially designed in the age of
waterfall development where projects developed from
requirements to delivery with very little
operational feedback.  Hence, a frequently asked
question about this work is the relevancy of that
$20^{th}$ century software management tool to
current practices.  That issue is the core
question addressed by this paper.  While there is
nothing inherently ``waterfall'' within the COCOMO
equations, COCOMO does assume that a size estimate
is available before the work starts.  Hence, it is
important to understand when changes to the size of
the software results in inaccurate COCOMO estimates.

Another complaint against the COCOMO equations is
that such ``model-based'' methods are less
acceptable to humans than ``expert-based methods''
were estimated are generated via committees of
experts. The advantage of such expert-based methods
is that if some new project has some important
feature that is not included in the COCOMO
equations, then human expertise can be used to
incorporate that feature into the estimate.
However,
such expert-based approaches have their limitations.
Valerdi~\cite{valerdi11} lists the cognitive biases
that can make an expert offer poor expert-based estimates.
Passos et al. offer specific examples for those
biases: they show that many commercial software
engineers generalize from their first few projects
for all future projects~\cite{passos11}.  J{\o}rgensen
\& Gruschke~\cite{jorgensen09} offer other results
consistent with Passos et al.  when they document
how commercial ``gurus'' rarely use lessons from
past projects to improve their future
expert-estimates.  More generally,
J{\o}rgensen~\cite{Jorgensen2004} reviews studies
comparing model- and expert- based estimation and
concludes that there there is no clear case that
expert-methods are better.  Finally, in 2015,
J{\o}rgensen further argued~\cite{jorg15} that
model-based methods are useful for learning the {\em
  uncertainty} about particular estimates; e.g.  by
running those models many times, each time applying
small mutations to the input data.

 One expert-based estimation method preferred by advocates of
 agile/devops is ``planning poker''~\cite{molokk08}
 where participants offer anonymous ``bids'' on the
 completion time for a project. If the bids are
 widely divergent, then the factors leading to that
 disagreement are elaborated and debated. This cycle
 of bid+discuss continues until a consensus has been
 reached.  Despite having numerous advocates,
 there are very few comparative studies of planning
 poker vs parametric methods. The only direct
 comparison we can find is the 2015 Garg and Gupta
 study that reports planning poker's estimates can be
 twice as bad as those generated via parametric
 methods~\cite{garg15}. One reason for the lack of
 comparisons is that COCOMO and planning poker
 usually address different problems: \bi
\item COCOMO is often used to negotiating resources prior to starting a project;
\item Planning poker is often used to adjust current activity within the resource allocation of a project.
  \ei
 Hence we say  there is no dispute between planning poker (that is an intra-project
task adjustment tool) and COCOMO (which is an pre-project tool for checking if enough resources are available
to start a project).

\section{Answers to Research Questions}
\label{sect:rqs}
\subsection{ RQ1: How big are real world ESP errors?}

In order to find real-world ESP errors,  we look to the historical record.
This study could find  two sources that mentioned ESP errors:
\bi
\item Source~\#1 covers fifty projects documented
by Jones \& Hardin~\cite{jones07a} from the U.S.
Department of Defense.
\item Source~\#2 covers 14 projects  from NASA. 
  \ei
  \fig{jones} describes the projects in Source~\#1.
  While the majority of
these were military-specific applications (command
and control functions), 17 of these are applications
types that could be seen in non-military domains.
 Only a minority of
these ($\frac{15}{50}=30\%$) projects were waterfall
projects where requirements were mostly frozen prior
to coding;
  For the remaining 35 of the Jones \& Hardin projects, there was ample opportunity
  for scope creep that could lead to inaccuracies in early life cycle estimates.

\begin{figure}[!t]
    \centering
  \scriptsize
  \begin{center}
      \begin{tabular}{|p{0.6in}|r|r|r|p{0.6in}|}\hline
      Complexity & N &  Product Line  & Environment & State of the Art\\\hline
Simple& 2& Existing& Existing &Current\\
Routine& 10& New& Existing& Current\\
Moderate& 14& New& New& Current\\
Difficult& 24& New& New& New\\\hline
   \end{tabular}

      \vspace{4mm}
      
  \begin{tabular}{|rp{1.8in}|rl|}\hline
        & & &Development\\
     N   &Application  types &N & processes\\\hline
     31 &Command and Control & 15 &Waterfall\\
    6 &Office Automation, Software Tools, Signals & 12 &Incremental\\
    5 &DFs, Diagnostic, Mission Plans, Sims, Utils; &  8 &Spiral;\\
   5 &Testing; & 15 &Undefined\\\cline{3-4}
   3 &Operating System &\\\cline{1-2}
  \end{tabular}
  \end{center}
  \caption{50 projects from~\cite{jones07a}.}\label{fig:jones}
\end{figure}
\begin{figure}
\centering
\includegraphics[width=3in]{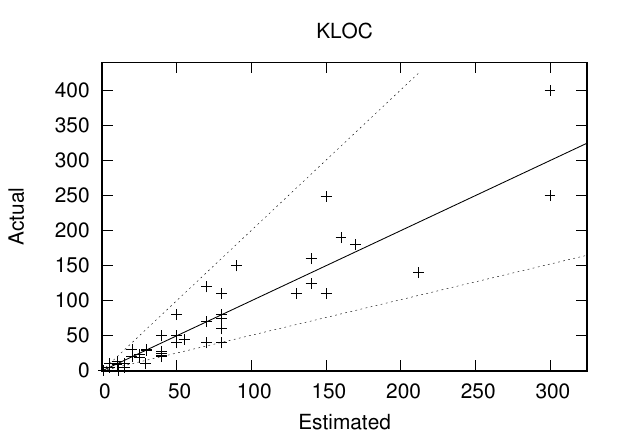}
\caption{Estimated and actual source lines of code from 50 projects~\cite{jones07a}.}\label{fig:ea}
\end{figure}

\fig{ea} shows the relationship between early life cycle estimates of KLOC vs final size for the 50
projects of Source~\#1. The diagonal line on that plot shows where the estimate equals the actual.
The dashed lines show the range within which the estimate is $\pm$~100\% of the actual.
The key observations from this Source~\#1 data are:
\bi
\item The estimates and actuals are often very similar.
  \item All the estimates fall within $\pm$~100\% of the actual.
    \ei
    
\begin{figure}
        \centering
      \scriptsize
      \begin{minipage}{.4\linewidth}
      \begin{tabular}{r|rr|}
    & pre-&pre-\\
   project & analysis&coding\\\hline
a&-44\%&-44\%\\
b&-13\%&-4\%\\
c&-6\%&-6\%\\
d&-4\%&-4\%\\
e&5\%&5\%\\
f&7\%&\\
g&10\%&10\%\\
h&54\%&54\%\\
i&64\%&64\%\\
j&69\%&69\%\\
k&78\%&78\%\\
l&95\%&52\%\\
m&206\%&10\%\\
n&236\%&236\%
      \end{tabular}\end{minipage} \begin{minipage}{.33\linewidth}
        \includegraphics[width=1.8in]{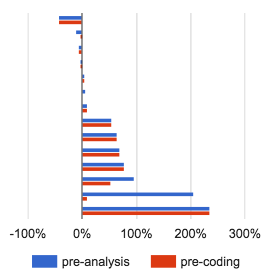}
        \end{minipage}
      \caption{Errors in estimates of final system size (measured in terms of KLOC)
        seen before analysis and coding in 14 NASA projects. Values in the left-hand-size table
        are shown graphically at right.
        All percentages here are percents on the final code size. For example, in the last
        line, Project N's final size was 236\% larger than predicted at the pre-analysis stage.
        Positive values denote initial under-estimates while negative values denote over-estimates
        (e.g. the size of the first four projects were initially over-estimated).}
        \label{fig:nasaloc}
    \end{figure}

  In order to check the external validity of these observations from Source \#1,
  we turn to the 14 NASA projects of Source~\#2. These projects cover the software used in the deep space missions of the
  Jet Propulsion Laboratory. Due to the highly innovative nature of this kind of
  software, this final size of this software could easily be incorrected
  estimated early in its life cycle.  Many of the details of those NASA systems
  is proprietary information so, in this article, we cannot describe the NASA
  systems at the same level of detail as Source~\#1. What  can be shown, in \fig{nasaloc},
  are the  ratios of actual/estimated KLOC values, where the estimates were generated
  prior to analysis and prior to coding. Usually, these two estimates were similar, but there
  are exceptions (e.g. the development effort estimate for project M was significantly
  adjusted after analysis).

 \begin{figure}[!b]

   \begin{center}
   \noindent{\scriptsize
\begin{tabular}{r|@{~}r|@{~}r|@{~}r|@{~}l}

Types of projects&\begin{sideways}COC81\end{sideways} & \begin{sideways}NASA93\end{sideways}& \begin{sideways}COC05\end{sideways} &\begin{sideways}NASA10\end{sideways}\\\hline 
Avionics&     &26&10&17\\\hline
Banking&       &      &13&     \\\hline
Buss.apps/databases&7&4&31&  \\\hline   
Control&9&18&13&     \\\hline
Human-machine interface&12&       &       &     \\\hline
Military, ground&       &       &8&     \\\hline
Misc&5&4&5&     \\\hline
Mission Planning&      &16&       &     \\\hline
SCI scientific application&16&21&11&     \\\hline
Support tools, &7&        &       &   \\\hline  
Systems&7&3&2&

%% NASA10& COC05 & NASA93& COC81 & Types of projects\\\hline\hline
%%      &     5 &4      &       & Misc\\\hline
%%      &       &        &  7   &  Support (tools, utilities, etc)\\\hline
%%      &      8&       &       & Military, ground\\\hline
%%      &      13&      &       & Banking\\\hline
%%      &      2&3      & 7     & Sys (OS, compilers, sensors,etc)\\\hline
%%      &      31&4     &  7    & Business apps/data processing\\\hline
%%      &       &16      &      & Mission Planning\\\hline
%%      &     13  &18     &   9 & Control\\\hline
%%      &       &       & 12    & Human=machine interaction\\\hline
%%      &      11& 21     & 16   & SCI scientific application\\\hline
%%  17  &     10 &26      &     & Avionics Monitoring
\end{tabular}}

\noindent\includegraphics[width=2.3in]{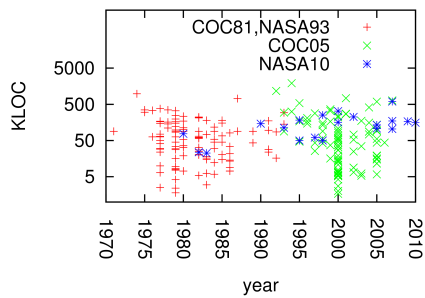}
\end{center}
\caption{Projects used by the learners in this study. \tab{cparems}
shows project attributes. 
COC81 is the original data from 1981 COCOMO book~\cite{boehm81}. 
This comes from projects dating 1970 to 1980.
NASA93 is NASA data collected  in the early 1990s
 about software that supported  the planning activities for the International
Space Station. 
Our other data sets are  NASA10 and COC05.
}\label{fig:types}
\end{figure}  

  The key observation from  this Source~\#2 data is that:
  \bi
\item The  $\pm$~100\% error seen in Source~\#1 covers all but one of the pre-coding
  NASA estimates. 
  \ei
  Hence, our first result is
  
\begin{lesson}
  Many real-world software projects usually have ESP errors of up to $\pm$~100\%.
\end{lesson}

%\begin{figure}
%\includegraphics[width=3in]{all.pdf}
%\caption{All errors seen in our two sources. 50th and 90th percentile on those errors are 91\% and 170\%.}\label{fig:allerr}
%\end{figure}

%% \fig{allerr} sorts all the {\em final/initial} ratios from our two sources. The horizontal lines in those figures show
%% the 50th and 90th value in that sort:
%% \bi
%% \item The median 50th percentile error is 91\%;
%%  \item The outlier 90th percentile error is 170\%.
%%    \ei
%%    Thus, based on \fig{allerr}, we can say that if perturb KLOC estimates by plus or minus 200\% then that more than covers the typical
%%    values seen in the historical record.

\subsection{RQ2: What is the impact of real-world ESP errors?}\label{sect:rq2}

This section reports the effort errors seen when
the size estimate in  real world project data are perturbed by
up to
$\pm$~100\%.

To conduct the perturbation study for identifying a point of tolerance, we use
COCOMO since its internal details have been fully
published~\cite{boehm00b}. Also, we can access a full implementation of the 2000
COCOMO model (see \fig{coc2}. Further, we have access to four interesting COCOMO data sets,
see \fig{types}.

The impact of noise in \textit{KLOC} is
studied by varying $\mathit{KLOC}$ as follows:
\begin{equation}
    \label{eq:kloc}
    \mathit{KLOC} = \mathit{KLOC}*((1- n) + (2*n*r))
\end{equation}
where $n \in [0.2, 0.4, 0.6, 0.8, 1.0]$ is noise level we are exploring and $r$ is a random number
$0 \le r \le 1$.

% \usepackage{multirow}
\begin{table}[!t]
\begin{center}

\scriptsize

\begin{tabular}{|r|c|c|c|c|}
\hline
\multicolumn{1}{|c|}{\multirow{2}{*}{\textbf{Name}}} & \multicolumn{2}{c|}{\textbf{Med}}     & \multicolumn{2}{c|}{\textbf{IQR}} \\ \cline{2-5} 
\multicolumn{1}{|c|}{}                               & \textbf{Rank} & \textbf{Med} & \textbf{Rank} & \textbf{Med} \\ \hline
    COCOMO2  & 1 & 43  & 1    & 1     \\ \cline{4-5}
20\%:COCOMO2 & 1 & 41   & 2    & 13   \\ \cline{4-5}
40\%:COCOMO2 & 1 & 41   & 3    & 28   \\ \cline{2-3}
60\%:COCOMO2 & 2 & 46   & 3    & 34   \\ 
80\%:COCOMO2 & 2 & 50  & 3    & 44  \\
100\%:COCOMO2 & 2 & 68  & 3    & 49    \\ \hline         
\end{tabular}
\vspace{-2mm}
\caption{NASA10}
\label{tab:nasa10}

\vspace{3mm}

\scriptsize
\begin{tabular}{|r|c|c|c|c|}
\hline
\multicolumn{1}{|c|}{\multirow{2}{*}{\textbf{Name}}} & \multicolumn{2}{c|}{\textbf{Med}}     & \multicolumn{2}{c|}{\textbf{IQR}} \\ \cline{2-5} 
\multicolumn{1}{|c|}{}                               & \textbf{Rank} & \textbf{Med} & \textbf{Rank} & \textbf{Med} \\ \hline
    COCOMO2  & 1 & 13   & 1    & 1     \\ \cline{4-5}
20\%:COCOMO2 & 1 & 14   & 2    & 8    \\ 
40\%:COCOMO2 & 1 & 19   & 2    & 14    \\ \cline{2-5}
60\%:COCOMO2 & 2 & 24   & 3    & 25   \\ 
80\%:COCOMO2 & 2 & 25   & 3    & 25   \\
100\%:COCOMO2 & 2 & 26  & 3    & 30     \\ \hline    
\end{tabular}
\vspace{-2mm}
\caption{COC05}
\label{tab:coc05}

\vspace{3mm}

\scriptsize
\begin{tabular}{|r|c|c|c|c|}
\hline
\multicolumn{1}{|c|}{\multirow{2}{*}{\textbf{Name}}} & \multicolumn{2}{c|}{\textbf{Med}}     & \multicolumn{2}{c|}{\textbf{IQR}} \\ \cline{2-5} 
\multicolumn{1}{|c|}{}                               & \textbf{Rank} & \textbf{Med} & \textbf{Rank} & \textbf{Med} \\ \hline
    COCOMO2  & 1 & 14   & 1    & 1     \\ \cline{4-5}
20\%:COCOMO2 & 1 & 14   & 2    & 5  \\ \cline{4-5}
40\%:COCOMO2 & 1 & 15   & 3    & 8   \\ \cline{4-5}
60\%:COCOMO2 & 1 & 16   & 4    & 12   \\ \cline{2-3}
80\%:COCOMO2 & 2 & 20   & 4    & 13   \\\cline{4-5}
100\%:COCOMO2 & 2 & 27  & 5    & 19    \\ \hline  
\end{tabular}
\vspace{-2mm}
\caption{NASA93}
\label{tab:nasa93}

\vspace{3mm}

\scriptsize
\begin{tabular}{|r|c|c|c|c|}
\hline
\multicolumn{1}{|c|}{\multirow{2}{*}{\textbf{Name}}} & \multicolumn{2}{c|}{\textbf{Med}}     & \multicolumn{2}{c|}{\textbf{IQR}} \\ \cline{2-5} 
\multicolumn{1}{|c|}{}                               & \textbf{Rank} & \textbf{Med} & \textbf{Rank} & \textbf{Med} \\ \hline
    COCOMO2  & 1 & 3   & 1    & 0     \\ \cline{4-5}
20\%:COCOMO2 & 1 & 4   & 2    & 2   \\ \cline{4-5}
40\%:COCOMO2 & 1 & 4   & 3    & 4   \\ \cline{2-5}
60\%:COCOMO2 & 2 & 6   & 4    & 6   \\ \cline{4-5}
80\%:COCOMO2 & 2 & 6   & 5    & 7   \\
100\%:COCOMO2 & 2 & 8  & 5    & 8    \\ \hline        
\end{tabular}
\vspace{-2mm}
\caption{COC81}
\label{tab:coc81}
\end{center}
\end{table}

% \begin{equation}\label{eq:mre}
% \mbox{$ \mathit{MRE}=\frac{abs(\mathit{actual} - \mathit{predicted})}{\mathit{actual}}$}
% \end{equation}

As per the advice of Shepperd and MacDonnell~\cite{shepperd12a},
we express effort estimation error as a ratio of some very simple method (in this
case, making a prediction by selecting at random some actual effort value from the training data).
Shepperd and MacDonnell's argument for this method is as follows: researchers
should report their methods as fractions showing how much their method improves over
some obvious baseline. Their preferred measure is the SA {\em standardized error}:

\begin{equation}\label{eq:sa}
\mbox{$ \mathit{SA}=\frac{abs(\mathit{actual} - \mathit{predicted})}{\mathit{\sum_{i=1}^{1000}|\pmb{choice}(all) - predicted|}/1000}$}
\end{equation}

\tab{nasa10}, \tab{coc05}, \tab{nasa93} \& \tab{coc81}
show the SA results
seen when all examples were passed to COCOMO-II. Since \eq{kloc} uses a random
number generator, we repeated that process 100 times. For all 100 repeated passes
through the data, SA was calculated with:
\bi
\item Estimated project size perturbed as per \eq{kloc};
\item $\mathit{actual}$ is the unperturbed value of the effort (taken from the data);

  \item  $\mathit{predicted}$ is the
    estimated value generated using the perturbed size estimate;
  \item
    $\mathit{all}$ is an array containing all the (not-perturbed) effort values in the dataset;
    \item
      $\pmb{choice}$ is a function that randomly picks one value from an array.
      \ei
      In \tab{nasa10}, \tab{coc05}, \tab{nasa93} \& \tab{coc81}:
      \bi
    \item
      The first column in each
table denotes the name of the method using the following nomenclature.
``$x$\%:COCOMO2'' represents COCOMO-II where KLOC
is perturbed with an error of $n$\% using \eq{kloc}.
\item Column 2 shows the ``rank'' of each result. In those figures,
  a row's rank increases if its SA results are significantly different than the row
  above. Note that for error measures like SA, {\em smaller} ranks are {\em better}.
  These ranks were computed using the Scott-Knott test of \fig{sk}. This test
  was adopted as per the recent recommendations of
  Mittas and  Angelis in IEEE TSE 2013~\cite{mittas13}.
\item
  Column 3 shows  the
  median and IQR for the median of the SA over the 100 repeats. The median value of a list
  is the 50th percentile value while the IQR is the 75th-25th percentile value.
  Note that for error measures like SA, {\em smaller} medians and smaller IQRs are {\em better}.
  \ei
  As might be expected, in these results,
  as ESP error increases, so to did the SA estimation error:
  \bi
\item For NASA10, from 43 to 68\%;
\item For COC05, from 13 to 26\%;
\item For NASA93, from 14 to 27\%;
\item For COC81, from 3 to 8\%.
  \ei
  That said, the size of the increase is surprisingly small. Even with errors up to 100\%:
  \bi
  \item
    The increased estimation error was sometimes very small: see the 5\% increase in COC81;
  \item
    The estimation error was never very large: i.e. never more than  the 25\% increase seen in COC81.
    \ei
\begin{figure*}[!ht]
    \centering\small
    \begin{minipage}[c]{\linewidth}
    \begin{mdframed}
    The Scott-Knott procedure is recommended by Mittas \& Angelis in their 2013 IEEE TSE paper~\cite{mittas13}.  This method sorts a list of $l$ treatments with $ls$ measurements by their median score. It then splits $l$ into sub-lists $m,n$ in order to maximize the expected value of differences  in the observed performances before and after divisions. E.g. for lists $l,m,n$ of size $ls,ms,ns$ where $l=m\cup n$:
     \[E(\Delta)=\frac{ms}{ls}abs(m.\mu - l.\mu)^2 + \frac{ns}{ls}abs(n.\mu - l.\mu)^2\]

    Scott-Knott then applies some statistical hypothesis test $H$ to check if $m,n$ are significantly different. If so, Scott-Knott then recurses on each division.Scott-Knott is better than an all-pairs hypothesis test of all methods; e.g. six treatments can be compared \mbox{$(6^2-6)/2=15$} ways.  A 95\% confidence test run for each comparison has  a very low total confidence: \mbox{$0.95^{15} = 46$}\%. To avoid an all-pairs comparison, Scott-Knott only calls on hypothesis tests {\em after} it has found splits that maximize the performance differences.

    For this study, our hypothesis test $H$ was a conjunction of the A12 effect size test(endorsed by Arcuri et al. in ICSE '11 \cite{arcuri11}) of  and non-parametric bootstrap sampling \cite{efron93}; i.e. our Scott-Knott divided the data if {\em both}
    bootstrapping and an effect size test agreed that the division was statistically significant (99\% confidence) and not a ``small'' effect ($A12 \ge 0.6$). 
    For a justification of the use of non-parametric
    bootstrapping, see Efron \&
    Tibshirani~\cite[p220-223]{efron93}.
    For a justification of the use of effect size tests
    see Shepperd \& MacDonell~\cite{shepperd12a}; Kampenes~\cite{kampenes07}; and
    Kocaguneli et al.~\cite{kocharm13}. These researchers
    warn that even if an
    hypothesis test declares two populations to be
    ``significantly'' different, then that result is
    misleading if the ``effect size'' is very small.
    Hence, to assess 
    the performance differences 
    we first must rule out small effects.
    Vargha and Delaney's
    non-parametric 
    A12 effect size test 
    explores
    two lists $M$ and $N$ of size $m$ and $n$:
    \[A12 = \left(\sum_{x\in M, y \in N} 
    \begin{cases} 
    1   & \mathit{if}\; x > y\\
    0.5 & \mathit{if}\; x == y
    \end{cases}\right) / (mn)
    \]
    This expression computes the probability that numbers in one sample are bigger than in another.
    This test was recently 
    endorsed by Arcuri and Briand
    at ICSE'11~\cite{arcuri11}.
    \end{mdframed}
    \caption{Scott-Knott Test}
    \label{fig:sk}
    \end{minipage}
\end{figure*}
Moreover, the median of the increased estimation error was usually smaller than
the inter-quartile range; e.g. for NASA10, the increase in the median error was
25\% while the inter-quartile range for 100\% perturbation was 59\%. Further, as
shown by the ``rank'' column in these results, all these results
were assigned the same value of ``rank''=1).
That is, while size estimate increases estimation error, those increases
{\em were not statistically significant}.

The reason for this result are clear: the variability associated effort estimates is not small.
\tab{nasa10}, \tab{coc05}, \tab{nasa93} \& \tab{coc81} report that the inter-quartile range in estimation error
can grow as large as 71\%. While size estimate error contributes somewhat to that error, it is clear
that  factors {\em other than size estimate error} control the estimate error.
These other factors are explored further in {\bf RQ3}. Meanwhile,
the clear result from {\bf RQ2} is:
\begin{lesson}
  In 265 real-world projects,
  ESP errors of up to $\pm$~100\% lead to estimate errors of only $\pm$~25\%.
\end{lesson}

This is surprising: in many real-world projects, large ESP errors
lead to  small estimation errors.
To explain this effect, we turned to our next research question.

\subsection{ RQ3:
  Within an effort estimation model,
  what is the maximum effect of making large changes to a size estimate?}

 Recall from the introduction that the exponential nature of the COCOMO equation
 made it seem as if COCOMO would be most susceptible to errors in lines of code.
 Yet we saw in the last section that COCOMO is remarkably {\em insensitive} to
 KLOC errors.  This section checks if that result is just some quirk of the 265
 projects studied above, or if it is a more fundamental property. 

 \eq{one} said  \mbox{$\mathit{effort} = \mathit{X*KLOC}^Y$}
 where \mbox{$Y={b + 0.01 \sum_i SF_i}$}.
 One explanation for the strange results of the last section is
 that the $Y$ coefficient on the exponential term are much smaller than the
 linear $X$ factors; i.e. the
 \begin{quote}
   {\em COCOMO effort estimate is
     effectively linear on $X$, and not exponential on $\mathit{KLOC^Y}$.}
 \end{quote}
 To prove this claim, we examine the coefficients of the terms in the COCOMO equation to
 see what effect changes in KLOC have on COCOMO's effort prediction
The coefficients learned by Boehm in 2000 for the
COCOMO were based on an analysis of 
161 projects from commercial, aerospace, government, and non-profit organizations~\cite{boehm00b}. At the time of that analysis,
those  projects   were of size 20 to 2000 KLOC (thousands of lines of code) and took between 100 to 10000 person months to build.
Boehm's   $SF_i$ coefficients
are presented in a table inside the front cover of the COCOMO-II text~\cite{boehm00a}(see \fig{coc2}).
When
  projects have ``very low'', ``low'', ``nominal'', ``high'', ``very high'' values in the COCOMO , then from that table it can be see that:
\begin{equation}\label{eq:sf1}
\begin{array}{r|l}
  &0.01 \sum_i  SF_i \\\hline
\mathit{very\; low} & 0.32\\
\mathit{  low} &   0.25\\
\mathit{nominal} &  0.192\\
\mathit{high} &  0.13\\
\mathit{very\; high} &  0.06  
\end{array}
\end{equation}
In 2000, Boehm proposed default values for $a,b$= $2.94,0.91$.
Those ranges of   where checked    by Baker~\cite{baker07} 
using 92 projects from NASA's Jet Propulsion Laboratory.  Recall from \eq{cocII}
that the $a,b$ local calibration parameters can be adjusted using local data. 
Baker checked those ranges by,  30 times, running the COCOMO
calibration procedure using 90\% of the JPL data (selected
at random). He reported that
 $a$ was approximately linearly related to $b$ as follows:
\[
\begin{array}{c}
\left(2.2 \le a \le 9.18\right) \bigwedge  \left(b(a,r) = -0.03a + 1.46 + r*0.1\right)
\end{array}
\]
Note that Baker's found ranges for $a$ included the $a=2.94$ value proposed by Bohem.

\begin{figure}[!t] 
\centering
\includegraphics[width=3.3in]{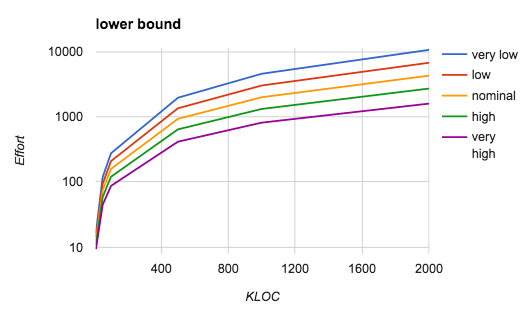}

\includegraphics[width=3.3in]{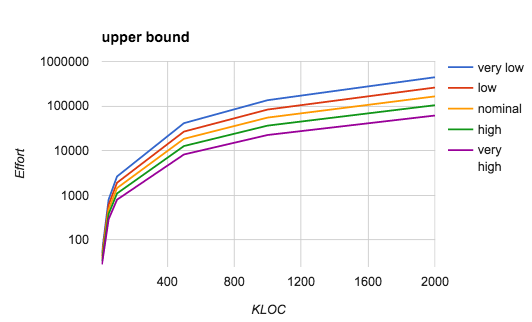} 
\caption{Growth in effort estimates as source code grows. Growth rate determined by \eq{sf3}. 
``Lower bound'' is the most optimistic projection (effort grows slowest as KLOC increases)
while ``Upper bound is most pessimistic.
For example, in `upper bound'' for ``very low'', $\mathit{effort} = \mathit{KLOC}^{1.71}$ (where 1.71 is the top-right figure of \eq{sf3}). }\label{fig:lowerupper}
\end{figure}

In the  above,  ``r'' is a random number $0 \le r \le 1$ so Baker's maximum and minimum $b$ values
were:
\[
\begin{array}{c}
b(2.2,\; 0) = 1.394\\
b(9.18,\; 1) =1.2846
\end{array}
\]
Combined with Boehm's default values for $b=0.91$, we say that in the historical record
there is evidence for $b$ ranging
\begin{equation}\label{eq:sf2}
0.91 \le b \le 1.394
\end{equation}
Combining the above with \eq{sf1}, we see that the $Y$ coefficient on the 
KLOC term in \eq{one} is 
\begin{equation}\label{eq:sf3} 
\begin{array}{r|l}
                  &  Y= b + 0.01 \sum_i SF_i \\\hline
\mathit{very\; low} &  1.22 \le Y \le 1.71\\
\mathit{  low} &  1.16 \le Y \le 1.65 \\
\mathit{nominal}& 1.10 \le Y \le 1.58    \\
\mathit{high} &  1.04 \le Y \le 1.52  \\
\mathit{very\; high} & 0.97 \le Y \le 1.46   
\end{array}
\end{equation} 
\fig{lowerupper} shows   $\mathit{effort} = \mathit{KLOC}^{\;Y}$ results using the coefficients
of \eq{sf3}. Note that the vertical axis of that chart a logarithmic scale.
On such a scale, an function that is exponential on the horizontal access will
appear as a straight line. All these plots bent over to the right; i.e. even
under the most pessimist  assumptions (see ``very low'' for ``upper bound''). That is:

\begin{lesson}
  In simulations over thousands of software projects,
  as KLOC increases, the resulting effort estimates increased {\em much less} than exponentially.
\end{lesson}

We show via the following analytical study 
that {\bf Result 3} can be explained with respect to internal structure
of the COCOMO parametric model. Using some algebraic manipulations of
our effort estimation model,
we can derive expressions from
 the (a)~the minimum and (b)~the
maximum possible effort estimate from this model.
By dividing these two expression, it is possible to create an fraction showing
the relative effect of changing size estimates, or any other estimates, within this model.

From \eq{cocII},
the minimum  
effort  is bounded by the  {\em sum} of the minimum scale factors
and the {\em product} of the minimum effort multipliers.
Similar expressions hold for the  maximum effort estimate. Hence,
for a given KLOC, the range of values is given by:
\[
0.057*\mathit{KLOC}^{\;0.97}  \le \mathit{effort} \le 115.6*\mathit{KLOC}^{\;1.71}\]
The exponents in the this expression come from \eq{sf3}. The linear terms come
from the product of the min/max effort multipliers from the 
 COCOMO-II text~\cite{boehm00b}.

Dividing the minimum and maximum values shows 
how    effort can vary for any given KLOC due to variations in the effort multipliers
and scale factors: 
\begin{equation}\label{eq:ration}
115.6/0.057 *\mathit{KLOC}^{\;1.71 - 0.97} = 2028*\mathit{KLOC}^{\;0.74}
\end{equation}
Note the large linear term (2028) coming from the effort multipliers and the
small exponential term (0.74) coming from the scale factors.
\eq{ration} shows that errors in the effort multipliers can change effort more than
errors in the size estimate.
Hence:

\begin{lesson}
  The net additional impact of ESP error is relatively small compared to the other sources of error associated within estimation models.
\end{lesson}

Hence, our conclusion is that
ESP errors can degrade project
effort estimates (see {\bf RQ2}). However, the size of that effect is much less than commonly feared.
Accordingly, we conclude that  ESP errors {\em are not} the dominant factor leading
to inaccurate effort estimates (see {\bf RQ3}).

\section{Validity}

\subsection{Sampling Bias}

The perturbation study used in {\bf RQ3} perturbed KLOC values according to the
ranges found in 64 projects discussed in {\bf RQ1}. Hence, the {\bf RQ3} conclusions 
are biases by that sample.

The other studies shown above were based on more data (265 projects).  While
that 265 does not cover the space of all projects, we note that it is a much
larger sample than what is seen in numerous other research papers in this arena.

While our sampling bias is clear, it also shows clearly how to refute our conclusions:
\bi
\item This study would need to be repeated in the ESP error in real world projects
  is observed to grow beyond $\pm$\;100\%.
  \ei

\subsection{External Validity}

One clear bias in this study is the use of the COCOMO model for studying the effects
of KLOC errors on estimates. Our case for using COCOMO was made above:
(a)~COCOMO s widely used in industry and government circles in the United States and China;
(b)~COCOMO's assumption that effort is exponentially proportional to KLOC seems to make it
exponentially sensitive to errors in KLOC estimates;
(c)~many aspects of COCOMO
 are shared by other models in widespread commercial
 use such as  SEER-SEM and Price-S (now called True S).

 As to external validity of our conclusions for  other effort estimation methods, Section 2.2
 offered arguments that ESP would have minimal impact on
 non-parametric models.

\section{Discussion}
If we accept the external validity of these conclusions, then the
next questions are:
\bi

\item
  What can be done to reduce the effort errors caused by all the non-size factors?
\item
  Why is
  software project development so insensitive to ESP errors?
  \ei
   As to the first question, we recommend feature selection. Elsewhere we have shown
  that as the number of training examples shrink, then it becomes more
  important to build models using just the few most important factors in the
  data. Automatic algorithms can find those most important factors~\cite{Chen:05}.
  
  As to the second question, it can be addressed via graph theory. While he
  never said it, we believe Boehm's core COCOMO assumption (that effort is
  exponentially proportional to size) comes for the mathematics of communication
  across networks. In an arbitrarily connected graph, a node has to
  co-ordinate with up to $2^{(N-1)}$ neighbors. If these nodes are software
  systems, then each such connection is one more team to co-ordinate with, or
  one more interaction to check in the test suite. Hence, a linear increase in a
  system size can lead to exponentially more complex co-ordination and testing.

  To avoid this extra exponential effort, it is necessary to reduce the number
  of other systems that interact with a particular node. This is the task of
  software architectures~\cite{Garlan:1994}.  A well-defined software
  architecture offers clear and succinct interfaces between different parts of a
  system~\cite{Parnas:1972}.  Either side of an interface, software may be very
  complex and inter-connected. However, given an interface that allows limited
  communication, the number of interactions between different parts of the
  systems are reduced. We conjecture that these reduced interactions are the core reasons
  why our models are not reporting effort being overly reactive to bad size predictions.

\section{Conclusion}

We have offered  evidence that we can be optimistic about
our ability to generate reasonably accurate early life cycle estimates, despite bad ESP:
\bi
\item
  From {\bf RQ1},
  we know that ESP errors seen in practice have limited
  ranges. Looking at \fig{ea} and \fig{nasaloc} we can see many projects where the estimated
  size was close to the actual final system.
  
\item
  In {\bf RQ2},
  we perturbed KLOC values within effort predictors
  (within the 
  maximum ranges found by {\bf RQ1}),
  The net effect of those perturbations was observed to
  be very small-- in fact statistically insignificant.
\item
  In {\bf RQ3} we checked the generality of the {\bf RQ2} conclusions.
  When we
compared the effects of KLOC error relative to errors in the other project
factors, we found that KLOC errors were  relatively less 
influential.
\ei
The last point is particularly significant. While there are many reasons why ESP
can fail (see the list of 5 points in the introduction), as shown above, the net
impact of those errors is relatively small.

In summary,
modern effort estimation models use much more than just size predictions.
While errors in size predictions increase estimate error, by a little,
it is important to consider all the attributes used by  effort model.
Future work should focus on how to better collect more accurate information about (e.g.)
the factors  shown in \tab{cparems}.  

\section*{Acknowledgments}
The work has partially funded by a National Science Foundation CISE CCF award \#1506586.

\small
\bibliographystyle{elsarticle-num}
% \balance
\bibliography{refs_min}  
\balance

\end{document}